\documentclass[aps, prb, twocolumn]{revtex4}
\usepackage{ams}
\usepackage{graphicx}

\usepackage{color}

\begin{document}

\title{Scaling Analysis and Application: Phase Diagram of Magnetic Nanorings and Elliptical Nanoparticles}
\author{Wen Zhang}
\email{zhangwen@usc.edu}
\author {Rohit Singh}
\author {Noah Bray-Ali}
\author {Stephan Haas}
\affiliation{Department of Physics and Astronomy, University of
Southern California, Los Angeles, CA 90089, USA}

\begin{abstract}
The magnetic properties of single-domain nanoparticles with
different geometric shapes, crystalline anisotropies and lattice
structures are investigated. A recently proposed scaling approach is
shown to be universal and in agreement with dimensional analysis
coupled with an assumption of {\em incomplete} self-similarity. It
is used to obtain phase diagrams of magnetic nanoparticles featuring
three competing configurations: in-plane and out-of-plane
ferromagnetism and vortex formation. The influence of the vortex
core on the scaling behavior and phase diagram is analyzed.
Three-dimensional phase diagrams are obtained for cylindrical
nanorings, depending on their height, outer and inner radius. The
triple points in these phase diagrams are shown to be in linear
relationship with the inner radius of the ring. Elliptically shaped
magnetic nanoparticles are also studied. A new parametrization for
double vortex configurations is proposed, and regions in the phase
diagram are identified where the double vortex is a stable ground
state.
\end{abstract}


\maketitle

\section{Introduction}
Magnetic thin films and nanoparticles have been intensively studied
during the last two decades \cite{De'Bell,Liu}, not only because of
their great potential for technological applications, but also
because of fundamental scientific interest. Many new phenomena come
about by imposing geometric restrictions in one \cite{De'Bell} or
more dimensions \cite{nanowire,Tsymbal,Cowburn,Ross}. It has been
demonstrated that nanoparticles show predominantly single domain
structure when their size is smaller than a characteristic length
scale \cite{Brown}. These single domain particles are promising
candidates for high density data storage \cite{Chou}, integrated
magnetic-electronic devices \cite{Parkin}, and applications in
biotechnology \cite{O'Grady}.

A great deal of attention has focused on arrays of magnetic
nanoparticles. In magnetic nanoparticle arrays, there are two
distinct issues of interest: the spin configuration of the
individual particles, and the interactions between them. Here we
focus on the first issue. The magnetic properties obtained under
this consideration are valid when the distance between the
individual particles is larger than twice the characteristic size of
the individual particles since it has been shown that the
interactions can be safely neglected under this condition.
\cite{Ross,Roshchin}

Within nanoparticles, different magnetic configurations have been
observed, including vortex, leaf, and flower states.\cite{Cowburn2}
Single-domain configurations have attracted continuous attention for
their obvious application potential. In particular, the magnetic
vortex, also known as non-localized soliton has been explored
recently for its application potential and interesting dynamics.
\cite{vortex1,vortex2,vortex3} In this work, we study magnetic phase
diagrams of such nanostructures as a function of their shape,
crystalline anisotropy and lattice structure.

On the numerical side, a scaling approach has been shown to be
effective in determining phase diagrams for cylinder
\cite{scalecylinder} and cone \cite{scalecone} shaped nanoparticles.
Here, we provide a systematic numerical study for different
geometric shapes. When the characteristic length scale is
sufficiently small, the shape of the particle is one of the dominant
factors determining its magnetic properties. Numerous experimental
investigations have addressed this and related issues of domain
structure. \cite{Cowburn,Cowburn3}

On the conceptual side, the scaling approach suggests a
self-similarity of magnetic nanoparticles.\cite{scaling_book} Ref.
\onlinecite{comment} tacitly assumes that magnetic nanoparticles
exhibit complete self-similarity with respect to the small parameter
$a/L_{ex}$, where, $a$ is the lattice spacing and $L_{ex}$ is the
magnetic exchange length. In fact, the particles exhibit only {\em
incomplete} self-similarity with respect to the lattice spacing in
certain circumstance, as we demonstrate in this work.  This
incomplete similarity agrees with dimensional analysis, as it must,
and with available numerical
data.\cite{scalecylinder,scalecone,scalemc}

The topology of the nanoparticle plays an important
role.\cite{Chien,nanoring1,nanoring2,ukraine_ring}  In a simply
connected topology, vortex states, for example, typically must have
a core region in which the spins point out of the vortex plane. In a
nanoring, furthermore, the inner radius $R_i$ provides an additional
length-scale with which to probe the self-similarity of magnetic
nanoparticles. We perform a scaling analysis, and show that
nanoparticles in this topology exhibit {\it complete} similarity
with respect to the lattice constant.  This is a consequence of the
additional length $R_i$ that plays the role of the lattice constant
in regulating the vortex core energy.

Quite a different scenario of the magnetization reversal was
revealed in elliptical particles.
\cite{ellipseNature,ellipseExp,ellipseExp2,Cowburn5} Numerical
simulations showed different spin configurations including multi
vortex states\cite{ellipseUsov2}. The double vortex configuration
confined in elliptically shaped ferromagnetic particle is especially
interesting, for it provides a model system for the study of static
and dynamic interaction between solitons (localized solution of
nonlinear equations)\cite{ellipseNature}. Many
efforts\cite{ellipseUsov1,ellipseUsov2} have been taken to obtain
the phase diagram of such systems, yet it is still an open problem.

The first topic of this paper is to reveal the essence of the
scaling approach and to verify its validity in terms of different
shapes, anisotropy and crystalline structure. Also, the effects of
these parameters on the phase diagrams are analyzed. The influence
of the vortex core on the scaling behavior and phase diagram is
investigated. Furthermore, the scaling approach is applied to
nanorings and elliptically shaped nanoparticles. The resulting phase
diagrams are given, and new and interesting phenomena are discussed.

\section{Model and Numerical Procedure}

In the absence of an external magnetic field, the Hamiltonian
($\mathcal{H}$)(or energy) of a magnetic nanoparticle consists of
three terms: exchange interaction, dipolar interaction, and
crystalline anisotropy. If each magnetic moment occupies a site of
the underlying lattice, $\mathcal{H}$ is given by

\begin{eqnarray}\label{sum}
\mathcal{H}&=&-J\sum_{<i,j>}\vec{S}_i\cdot\vec{S}_j \nonumber \\ & &
+D\sum_{i,j}\frac{\vec{S}_i\cdot\vec{S}_j-3(\vec{S}_i\cdot
\hat{r}_{ij})(\vec{S}_j\cdot \hat{r}_{ij})}{r_{ij}^{3}} +U_k,
\end{eqnarray}

where $J>0$ is the ferromagnetic exchange constant (or exchange
integral, measured in units of energy)\cite{Kittel}, which is
assumed to be non-zero only for nearest neighbors (nn), $D$ is the
dipolar coupling parameter and $\vec{r}_{ij}$ the displacement
vector between sites $i$ and $j$. The anisotropy term $U_k$ can take
various forms\cite{Kittel} among which the most common are uniaxial
anisotropy $U_k=K\sum_i sin^2\theta_i$ , where $\theta_i$ is the
angle $\vec{S}_i$ makes with the easy axis, and cubic anisotropy
$U_k=K\sum_{i}[\alpha_i^2 \beta_i^2+\beta_i^2 \gamma_i^2+\alpha_i^2
\gamma_i^2]$, where {$\alpha_i,\beta_i,\gamma_i$} are the direction
cosines of $\vec{S}_i$. Note that $K$ is the single site anisotropy
energy (not an energy density). For most materials, the
dimensionless ratio $D/Ja^3$ falls in the range of $10^{-3}$ and
$10^{-4}$, where, the lattice constant $a$ is approximately 3\AA.
The dimensionless ratio $Ka^3/D$ lies between 0 and 10. We choose
$Ka^3/D=1$, $Ja^3/D=5000$ and $a=3\AA$ in the following
calculations, unless they are specified otherwise.

The objects studied in this paper are magnetic nanoparticles with
various shapes and anisotropies. In such systems, three dominant
competing configurations have been identified\cite{scalecylinder}:
(I) out-of-plane ferromagnetism with the magnetization aligned
parallel to the nanodot base; (II) in-plane ferromagnetism with the
magnetization perpendicular to the base; (III) a vortex state with
the magnetic moments circling in the base plane. Double vortex
states in elliptically shaped particles will be discussed in detail
later. A typical phase diagram for a cylinder is shown in Fig. 1,
which exhibits these three phases as a function of the cylinder
radius $R$ and its height $H$. Note that there can be other
metastable configurations, such as the buckle
state\cite{Cowburn4,Metlov}, which are not considered here. These
states result from the competition between the exchange and dipolar
interaction. The exchange interaction tends to align spins in the
same direction, whereas the dipolar interaction encourages spins to
minimize their magnetostatic energy resulting in the shape
anisotropy. Thus spins align in-plane in a flat disk while they
point out-of-plane in an elongated cylinder. The vortex state is
also a result of dipolar interactions since it nearly eliminates the
demagnetization field.

In order to obtain phase diagrams such as the one shown in Fig. 1,
one could resort to analytical calculation based on a continuum
model. However this approach is limited to highly symmetric shapes
and magnetization configurations. An alternative is to use numerical
simulations. These can be powerful and universal but are often
limited by computational resource. As outlined in Ref.
\onlinecite{scalecylinder}, the major technical problem is that the
number of magnetic moments in systems of physical interest is of the
order of $10^9$, which cannot presently be handled, even by high-end
supercomputer facilities. To overcome this restriction, a scaling
approach was recently proposed and demonstrated for cylinder
\cite{scalecylinder} and cone \cite{scalecone} shaped nanoparticles.
They showed that the phase diagram for an artificial small $J'=xJ$
($x<1$) could be scaled to the phase diagram for the original $J$
according to $L'=x^\eta L$ ($\eta \simeq 0.55$ and $L$ can be
$R,H$). The phase boundary for small $J'$ appears at small sizes
which involve less number of spins so that lots of computing time is
saved.

\begin{figure}[h]
\includegraphics[height=.25\textheight]{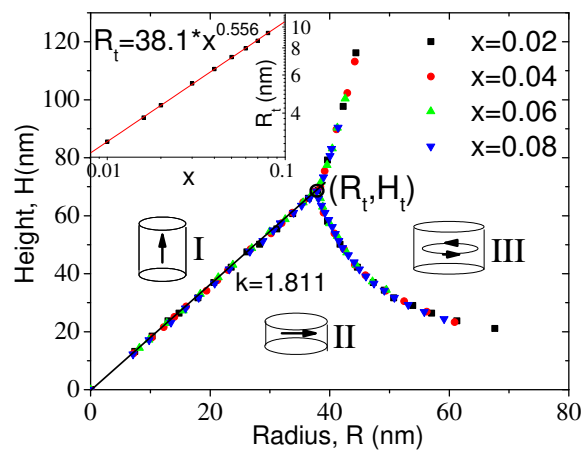}
\caption{(color online) Scaled phase diagram of cylindrical magnetic
nanoparticle ($Ka^3/D=1$ and $Ja^3/D=5000$) as a function of its
radius and height. The underlying lattice is simple cubic with
$a=0.3\AA$. The three competing phases are (I) out-of-plane
ferromagnetism, (II) in-plane ferromagnetism, and (III) the vortex
state. The transition lines are obtained according to the scaling
approach discussed in the text. The inset shows the dependence of
the radius at the triple point on the scaling factor x.}
\end{figure}

This proposal is equivalent to dimensional analysis coupled with a
statement of {\em incomplete} similarity.  We seek, for example, to
find the height $H$ separating the vortex phase from the
ferromagnetic phase(s).  In addition to the two governing parameters
$J,D$ appearing explicitly in Eq.~\ref{sum}, we also have the
radius, $R$ of the cylinder and the lattice constant $a$.  Thus we
seek a physical law for the critical height of the following form:
\begin{equation}
H=f(J,D,R,a). \label{physical_law}
\end{equation}

From dimensional analysis, only two of the four governing parameters
have independent dimensions.  Following convention, we choose the
independent parameters to be $J$ and $D$, define the exchange length
$L_{ex}=a\sqrt{Ja^3/D}$, and express the scaling law in
dimensionless form:
\begin{equation}
\Pi=\Phi(\Pi_1,\Pi_2),
\end{equation}
where $\Pi=H/L_{ex}$, $\Pi_1 = R/L_{ex},\Pi_2=a/L_{ex}$, and the
scaling function $\Phi$ does not depend on the governing parameters
of independent dimension $J,D$.



Now the typical values $a\sim0.3$nm, $L_{ex}\sim20$nm give $\Pi_2\ll
1$.  We are tempted to suggest complete similarity with respect to
the small, dimensionless governing parameter
$\Pi_2$.\cite{scaling_book} Hence we consider the limit $\Pi_2=0$:
\begin{eqnarray}
\Pi&=&\Phi(\Pi_1,0)\equiv \Phi_1(\Pi_1), \label{pizero}
\end{eqnarray}
where $\Phi_1$ is independent of $J$ and $D$ and $a$. Recasting in
original variables, we have that $H=L_{ex}\Phi_1(R/L_{ex})$. Notice
now the invariance of this relation under the following rescaling of
the governing parameters and critical height:
\begin{eqnarray}
 J'&=&x J, \nonumber \\
 D'&=&D \nonumber \\
 R'&=&x^{1/2} R \nonumber \\
 a'&=& a \nonumber \\
 H'&=& x^{1/2} H,
\label{transform}
 \end{eqnarray}
where, $x$ is any positive number.  One way to see this is to notice
that all lengths entering Eq.~\ref{pizero} get rescaled by the same
amount $x^{1/2}$, thus the dimensionless ratios $\Pi,\Pi_1,$ are
invariant. The invariance of Eq. \ref{physical_law} under this
transformation is a consequence of dimensional analysis combined
with complete similarity with respect to the dimensionless governing
parameter $\Pi_2=a/L_{ex}$.

Interestingly, the numerical calculations under the assumption of a
core-free vortex phase, where only the magnetic moment located
exactly at the center of the vortex has a component pointing out of
the vortex plane, do not obey this
scaling.\cite{scalecylinder,scalecone}  Instead, they exhibit only
incomplete similarity with respect to
$\Pi_2=a/L_{ex}$.\cite{scaling_book} Namely, for small values of
$\Pi_2\ll 1$, we have
\begin{eqnarray}
\Pi=\Pi_2^{1-2\eta}\Phi_2(\frac{\Pi_1}{\Pi_2^{1-2\eta}}),
\label{pieta}
\end{eqnarray}
where, the constant $\eta\approx 0.55$ does {\it not} follow from
dimensional analysis.  Here, $\Phi_2$ is independent of $J,D,$ and
$a$.  The special case of complete similarity is obtained when
$\eta=1/2$.  This implies that the physical law Eq.
\ref{physical_law}, is {\it not} invariant under the transformation
in Eq. \ref{transform}.  Rather, we find invariance under the
modified transformation:

\begin{eqnarray}
 J'&=&x J, \nonumber \\
 D'&=&D \nonumber \\
 R'&=&x^\eta R \nonumber \\
 a'&=& a \nonumber \\
 H'&=& x^\eta H.
\label{transformeta}
 \end{eqnarray}

This is precisely the transformation described in Ref.
\onlinecite{scalecylinder}, and is a consequence of dimensional
analysis combined with incomplete similarity with respect to the
small dimensionless parameter $\Pi_2=a/L_{ex}.$ This {\em
incomplete} similarity results from the fact that there is a
singularity in the magnetization function.

Whenever we are presented with a scaling phenomenon such as Eq.
\ref{pieta}, we have the opportunity to save considerable
computational and experimental effort.  The scaling law expresses a
physical similarity between systems with different values of the
governing parameters, so that we can use one to study the other.  In
particular, the authors of Ref. \onlinecite{scalecylinder} suggest
that we study small systems with small exchange constant $J$, which
are less computationally intensive to simulate.  Then use Eq.
\ref{transformeta} to scale up the results to the large systems with
large exchange constant that are of immediate physical and
technological interest.   The proposal is justified by the
incomplete similarity of the physical law Eq. \ref{physical_law}
with respect to the small, dimensionless parameter $\Pi_2=a/L_{ex}$.
Other physical quantities of nanomagnets may satisfy incomplete
similarity, including dynamic and thermal properties.\cite{scale,
scalemc,scale2}.

\section{Results and Discussion}

\subsection{Shape, anisotropy, and lattice structure}

{\em Incomplete} similarity with respect to the lattice constant
occurs in magnetic nanoparticles regardless of cross-sectional
geometry, crystalline anisotropy, or lattice structure. To
illustrate the use of the scaling procedure, let us first consider
the example of a cylindrical nanoparticle. Using the 2000-node 15.78
teraflop high-performance supercomputer at University of Southern
California (USC), the energies of the competing phases were
evaluated throughout the parameter plane spanned by the cylinder
radius $R$ and height $H$ for systems with up to 400,000 sites. The
scaling procedure was then used to collapse the resulting phase
diagrams with different scaling factors, four of them (x=0.02, 0.04,
0.06, and 0.08) given in Fig. 1 as examples. Note that there is a
triple point $(R_t,H_t)$, which is used to extract the scaling
exponent, shown in the inset of Fig. 1. For the sake of simplicity,
a simple cubic underlying lattice structure with cubic crystalline
anisotropy and the ``core-free" vortex state is adopted. Discussion
about other structures and the effect of the core will come later.

The scaling exponent $\eta=0.556$ is consistent with the previous
result \cite{scalecylinder}, suggesting incomplete similarity with
respect to the lattice constant in this case. It is observed that
the slope of the line separating the two ferromagnetic phases is
k=1.811, which is in exact agreement with the analytical solution
given previously \cite{Aharoni} and argued later \cite{comment}.

Since an enormously wide range of magnetic properties can be
obtained by using different geometric shapes\cite{Cowburn}, it is of
great interest to see whether nanoparticles with different
cross-sectional geometry exhibit incomplete similarity as well. To
answer this question, here we consider prism shaped nanoparticles
with triangular, square, pentagonal, and hexagonal cross sections.
From the results shown in Fig. 2(a) we find that within an error bar
of $2\%$, these different geometries have the same scaling exponent
showing incomplete similarity. In spite of the apparently universal
scaling behavior, it is also evident that different geometries do
favor different spin configurations. More precisely, the more
symmetric the cross section is, the more the vortex phase is
favored. Obviously, cylindrical nanodot favors the vortex
configuration the most. Another property of interest is the slope
$k$ of the line separating the two ferromagnetic (FM) phases. Fig.
2(b) shows this slope as a function of the cross section area. To
compare the various polygon shapes, they have been normalized such
that the distance from the corner of each polygon to its center is
unity. The slope is found to increase with the basal area. This
trend is easy to understand, since the two FM configurations are
determined by dipolar interactions, i.e. via the demagnetizing field
which in turn is related to the surface area. Quantitatively the
slope is expected to be approximately proportional to the square
root of the area, which is found to be in agreement with the
numerical results shown in Fig. 2(b).

\begin{figure}[h]
\includegraphics[height=.25\textheight]{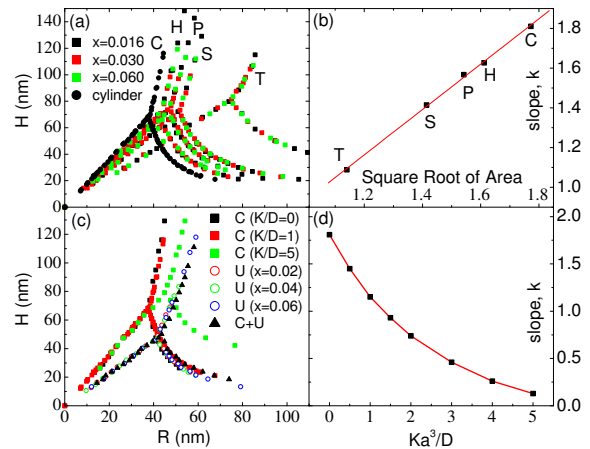}
\caption{(color online) (a) Scaled phase diagrams for prism shaped
nanoparticles. The radii $R$ are defined as the distance from the
base center to the corner of the polygons. The extracted scaling
exponents for the triangle (T), the square (S), the pentagon (P) and
the hexagon (H) are 0.556 (T), 0.557 (S), 0.563 (P) and 0.559 (H)
respectively. (b) The slope (k) of line separating phase I and II
versus the square root of the cross section area of nanodot with
unit radius. (c) Phase diagrams of cylindrical nanoparticles with
different anisotropies. Solid squares represent cubic anisotropy (C)
of different magnitude. Open circles with different colors represent
uniaxial anisotropy(U) with $Ka^3/D=1$ showing valid scaling
behavior with $\eta=0.56$. Solid triangles represent combination of
both anisotropies (U+C) with $Ka^3/D=1$. (d) The slope k versus the
strength of the uniaxial anisotropy. }
\end{figure}

In the following analysis of the universality of scaling for various
crystalline anisotropy and underlying lattice structures, we will
focus on cylindrical shapes for the simple reason that these are
most commonly found in the existing experimental literature. Fig.
2(c) gives phase diagrams for different anisotropies. In accordance
with intuition, cubic anisotropy favors the two ferromagnetic phases
equally, i.e. the slope separating these two phases does not depend
on $Ka^3/D$, and at the same time suppresses vortex formation.
Hence, one should consider materials with {\em{small}} cubic
anisotropy if one wishes to stabilize the vortex state. Besides
cubic anisotropy, another prevalent type is the uniaxial anisotropy.
This anisotropy typically exists in hexagonal close-packed (hcp)
lattices, but it can also occur in cubic lattices due to coupling to
the substrate or other parts of the environment. In our calculation,
the easy axis is set to be along the axis of the cylinder. The
resulting phase diagram is shown in Fig. 2(c). We observe that
uniaxial anisotropy does not affect the scaling behavior and
exponent. However, a feature worth mentioning is that uniaxial
anisotropy does change the slope of the line separating the two
ferromagnetic phases, favoring out-of-plane alignment (phase I). The
larger the value of $Ka^3/D$, the smaller the slope (see Fig. 2(d)).
Meanwhile, when both anisotropies are present, the slope is
dominated by the uniaxial term. Hence an analysis of this slope can
be used to determine the uniaxial anisotropy experimentally, based
on the information given in Fig. 2(d).

Various lattice structures exists in nature. It is important to know
whether the scaling technique depends on lattice structure. We
calculated the phase diagram for hcp and face centered cubic (fcc)
lattices and their variance by rotating the lattice structure in the
cylinder. The results remain invariant as long as all parameters
($Ja^3/D$, $Ka^3/D$ and density of spins) are kept the same and x is
not too small. The above results indicate that the scaling behavior
is robust to details of lattice structure, crystalline anisotropy,
and geometric shape.

\subsection{Particles with Core Structure}
\begin{figure}[h]
\includegraphics[height=.25\textheight]{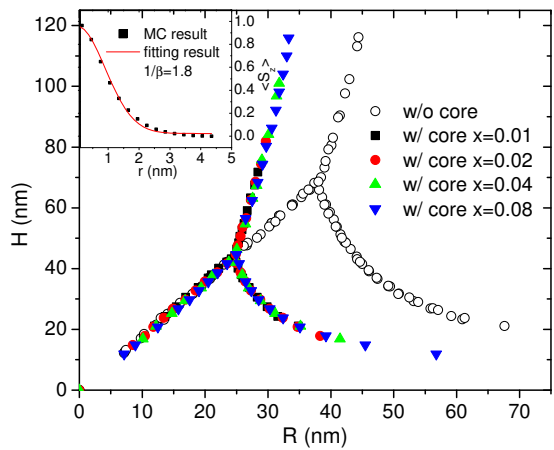}
\caption{(color online) Scaled phase diagram of a single-domain
cylindrical magnetic nanoparticle taking the vortex core into
consideration. The black hollow circles represent the phase diagram
for cylindrical nanoparticle with core free model taken from Fig. 1.
The scaling exponent $\eta=0.5$. The inset shows the fitting of the
core function to the MC result for the case of $J'/D=100$.}
\label{complete}
\end{figure}
Interestingly, magnetic nanoparticles with core structure exhibit
{\it complete} similarity with respect to the lattice constant (See
Fig. \ref{complete}). Similar effects have been reported by Landeros
et. al.\cite{Vargas}  To analyze the effect of the core, we choose
an ansatz ($S_z= \exp (-2r^2\beta^2$)) introduced by Feldtkeller and
Thomas\cite{core}. We fit the results of Monte Carlo (MC)
simulations with this ansatz and obtain acceptable agreement (see
the inset of Fig. 3). From dimensional analysis, the core size
$1/\beta$ obeys a scaling law of the following form:
\begin{equation}
1/\beta=L_{ex}\Phi_\beta(R/L_{ex},H/L_{ex},a/L_{ex}),
\end{equation}
where $L_{ex}$ is the magnetic exchange length, as before, and
$\Phi_\beta$ is a scaling function, independent of $J,D$.
Numerically, we find that the scaling function $\Phi_\beta$ is
approximately independent of all its arguments, giving roughly $
1/\beta\approx0.6 L_{ex}$. We use this as an additional governing
parameter in the numerical calculations.

In the presence of the core, the critical height now satisfies a physical law of the form:
\begin{equation}
H=g(J,D,R,a,1/\beta). \label{physical_law2}
\end{equation}
From dimensional analysis, we find again only two independent
governing parameters, define $L_{ex}$, and write:
\begin{equation}
H=L_{ex} \Phi_g(R/L_{ex},a/L_{ex},1/(\beta L_{ex})).
\label{scalephig}
\end{equation}
Numerically, we find that $\Phi_g$ approaches a constant as its
second argument $a/L_{ex}$ becomes small. This is evidenced by the
collapse of the phase diagrams in Fig. 3 with $\eta=1/2$. The
collapse implies invariance under the transformation in Eq.
\ref{transform}, and thereby the complete similarity with respect to
the lattice constant. However, this is consistent with the
incomplete similarity with respect to an exhibited in the core-free
approach, since we have an additional dimensional length $1/\beta$
that plays the role of the lattice constant. Similar results obtain
when we change the topology of the nanoparticle and introduce an
inner radius.

As to the phase diagram itself, the core stabilizes the vortex
configuration significantly, pushing the phase boundary between FM
and the vortex phase to smaller values of $R$ and $H$ by about $35
\%$. Similar effects would affect Fig. 2(a)(c) as well.

\subsection{Cylindrical Nanorings}

Next we consider the effects of changes in topology on the phase
diagram. More precisely, we investigate the phase diagram of hollow
cylinders, i.e. nanoring structures characterized by an inner radius
$R_i$, an outer radius $R$, and a height $H$. We find, as in the
previous section, that the critical height exhibits {\it complete}
similarity with respect to the lattice constant ($\eta\approx 1/2$).
This is a consequence of the additional length $R_i$ that plays the
role of the lattice constant in regulating the vortex core energy.

\begin{figure}[h]
\includegraphics[height=.25\textheight]{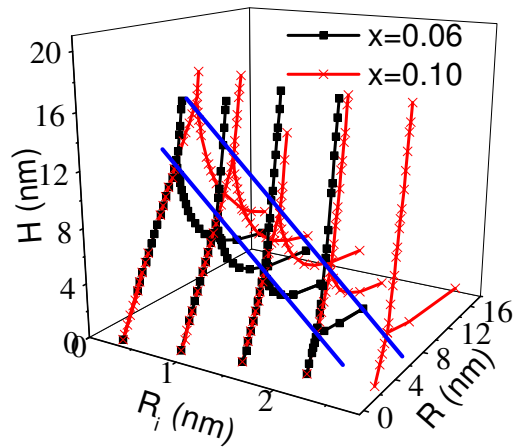}
\caption{(color online) Phase diagrams of a cylindrical nanoring for
two different x. There are two competing ferromagnetic phases at
small ($R,H$) and a vortex phase at large ($R,H$). Because of the
finite inner radius $R_i$, the onset of the phase transition line
between the two ferromagnetic phases is shifted to finite values of
$R$. Also, the vortex regime is more extended for larger $R_i$. The
blue lines in the figure are guides to the eye, indicating that the
triple points form approximately a straight line. }\label{3dring}
\end{figure}

\begin{figure}[h]
\includegraphics[height=.25\textheight]{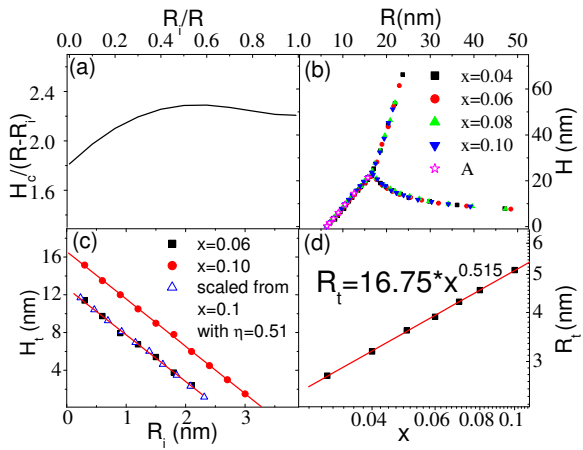}
\caption{(color online) (a) For cylindrical nanorings, the phase
transition line $H_c(R,R_i)$ separating the two ferromagnetic
regimes is not straight, in contrast to the topologically connected
objects discussed above. (b) Phase diagram for a cylindrical
nanoring. ($R_i=6.3nm$, $J/D=5000$). The data ``A" represent
analytical phase transition lines calculated from (a), which are
observed to coincide with the numerical results. (c) Height at the
triple point ($H_t$) versus inner radius ($R_i$). The best fit for
x=0.1 is $H_t = 16.5 - 5(\pm 0.03) \times R_i$, whereas the best fit
for x=0.06 is $H_t = 12.7 - 4.95 (\pm 0.09)\times R_i$. Hence, the
two lines are approximately parallel, and can thus be collapsed via
scaling with $\eta=0.51$. (d) The triple point radius ($R_t$) versus
x.  }\label{ring}
\end{figure}

Fig. \ref{3dring} shows three-dimensional phase diagrams in the
$(R_i,R,H)$ parameter manifold of the nanoring topology for two
different values of the exchange couplings $J'$. Again, one observes
two ferromagnetic regimes at small ($R,H$) values, competing with a
vortex phase at larger ($R,H$). Moreover, one finds that for larger
inner radii $R_i$ the the vortex phase is more extended. This
confirms the idea that the ring structure stabilizes the vortex
configuration. The reason for this is that the core area, which
typically pays a high energy penalty, is deliberately avoided in the
ring structure. Another new feature of these phase diagrams is that
the line separating the two ferromagnetic phases is not straight
anymore. Instead, it now starts at finite $R=R_i$, and its slope
changes smoothly to ~1.81 as the ratio between $R$ and $R_i$ becomes
very large. This relationship can be observed clearly in Fig.
\ref{ring}(a) which is derived from Eq.(11,13) in Ref.
\onlinecite{nanoring1}. Here we calculate the relationship between
the critical height $H_c (R,R_i)$ as a function of $(R-R_i)$,
resulting in the ``star" symbols in Fig. \ref{ring}(b) which align
exactly with the line of our numerical calculation. Finally, the
most surprising feature of the phase diagrams in Fig. \ref{3dring}
is that the triple points $(R_t, H_t)$ for different $R_i$
approximately form a straight line indicated by the two blue lines.
This property is shown more clearly in Fig. \ref{ring}(c), i.e.
cylinder height at the triple point ($H_t$) versus $R_i$. It gives
us a critical $R_{ic}$ beyond which there exists no in-plane
ferromagnetic phase.

Another observation worth mentioning is that the intersect phase
diagram in the $R_i=0$ plane of the cylindrical nanoring does not
coincide exactly with the phase diagram of the simply connected
cylinder (Fig. 1). It is closer to the case when the core structure
is considered (Fig. \ref{complete}). This phenomenon happens for the
scaling exponent as well, which will be discussed right below.

One last feature to be discussed here is that the line connecting
$H_t$ (see Fig. \ref{3dring} and Fig. \ref{ring}(c)) is parallel for
different values of the exchange coupling $J'$. Comparing the two
phase diagrams for different $J'$, we anticipate that there exists
scaling behavior here as well, as long as all three coordinates
$(R_i, R, H)$ are scaled. However, some difficulties arise since
$R_i$ should be different for different $J'$s, meaning that one
would need to know the scaling exponent $\eta$ in advance. Luckily,
we can estimate the value of $\eta$ from Fig. \ref{ring}(c) as the
two straight lines should scale if there is a scaling behavior. Thus
we first attempt to scale these two lines and find that they fit
best when $\eta\simeq0.51$. Then we use this $\eta$ to scale $R_i$
and attempt to see whether the scaling behavior holds. Fig.
\ref{ring}(d) shows the result. The scaling exponent is $\eta=0.515$
which is within 1\% of the estimated value 0.51. It is much closer
to 0.5 in the finite core case, implying complete self-similarity,
since an additional length $R_i$ is added and neglecting the core
structure in the vortex state has little effect for ring structure.
With these results, we can easily calculate the critical inner
radius $R_{ic}\simeq11nm$ for the parameters we choose, above which
a flat nanoring is always in the vortex phase. This is quite small
compared to typical nanorings fabricated experimentally\cite{Chien},
and suggests that nanorings are generically in the vortex phase,
since they are typically flat with height small compared with the
width.

\subsection{Elliptically Shaped Particles}

It has recently been observed that there exists a double vortex
configuration in elliptically shaped ferromagnetic
particles.\cite{ellipseUsov1,ellipseUsov2,ellipseExp,ellipseExp2}
The full phase diagram for this case as a function of height,
semi-major axis ($R_a$) and semi-minor axis ($R_b$), however, has
not yet been calculated. One of the difficulties to determine this
phase diagram, using the technique outlined above, lies in finding
an adequate parametrization of the double vortex state. The naive
approximation of two single vortices is far from satisfying (see
Fig. 6(c)). As we will see below, the energy of two single vortices
with discontinuous magnetization along the minor axis is
significantly higher than of a true double vortex with continuously
varying magnetization (see Fig. 6(a)). Without an accurate
parametrization of the double vortex one could only rely on Monte
Carlo or micromagnetic simulations which are extremely time
consuming, and this would make it impossible to obtain a complete
phase diagram.
\begin{figure}[h]
\includegraphics[height=.45\textheight]{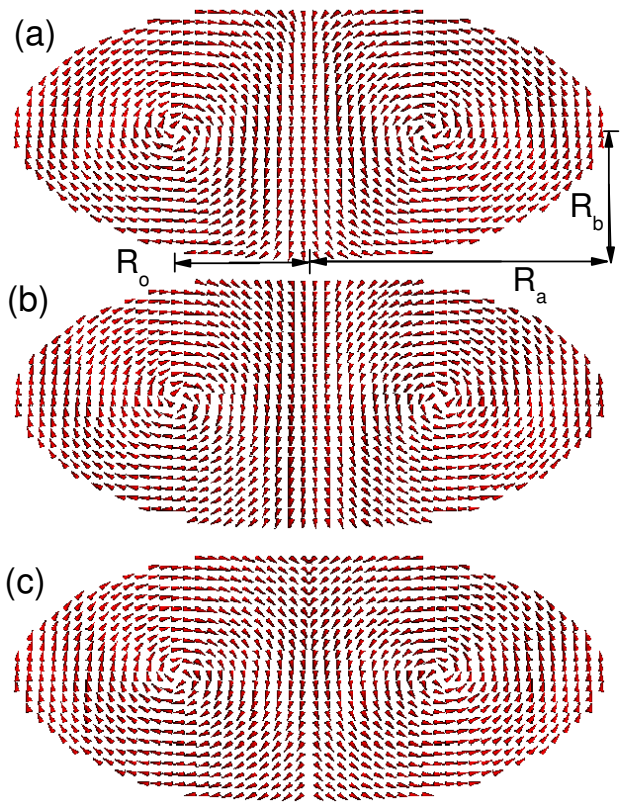}
\caption{(color online) Double vortex configuration for
$J/D=10$($x=0.002$), $R_a/R_b=2$ (arrows represent the directions of
magnetization). (a) Monte Carlo simulation result, $ea^3/D=21.12$
(b) Our parametrization, $ea^3/D=21.11$, $\mathcal{F}$ between b and
a is 0.990 (c) naive parametrization (two single vortices),
$ea^3/D=20.91$, $\mathcal{F}$ between c and a is 0.974. $e$ is the
energy per spin and $\mathcal{F}$ is the fidelity defined in the
text. }
\end{figure}

Here we propose a simple function to parametrize the double vortex.
In our Monte Carlo simulations, we observe that the shape of the
double vortex (Fig. 6(a)) looks much like the equipotential lines of
two electric point charges with opposite signs placed at the centers
of the vortex cores (Fig. 6(b)). By symmetry these cores should lie
on the major axis of the ellipse. Let the distance from the core
centers to the center of the ellipse be $R_o$. Then the vector field
$\vec{S}(\vec{r})$ is given by

\begin{equation}
 \vec{S}(x,y)=\frac{-E_y\hat{i}+E_x\hat{j}}{E^2_x+E^2_y}
 \label{ellipse}
  \end{equation}

where,
\begin{eqnarray}\label{xy}
E_x&=&\frac{x-R_o}{[(x-R_o)^2+y^2]^{3/2}}-\frac{x+R_o}{[(x+R_o)^2+y^2]^{3/2}}\nonumber\\
E_y&=&\frac{y}{[(x-R_o)^2+y^2]^{3/2}}-\frac{y}{[(x+R_o)^2+y^2]^{3/2}}\nonumber
\end{eqnarray}

Interestingly, the optimal positions of the vortex cores yielding
the lowest energy configurations do not coincide with the ellipse
foci, but are located at non-trivial positions on the major axis
with constant $\kappa=R_o/R_a$. $\kappa$ depends almost exclusively
the aspect ratio ($R_a/R_b$), and depends only very weakly on size.
Within the range we examined ($H<40nm,R_a<30nm$), $\kappa$ decreases
by only $2\%$ as the size is increased. For different aspect ratios
we find $\kappa=0.44\pm0.1$. These values coincide with recent
experimental results\cite{ellipseExp,ellipseExp2,ellipseNature}. We
choose $R_a/R_b=2$ as an example. In this case, $\kappa=0.44$. To
quantify the quality of our parametrization of the double vortex, we
look at the energy per spin ($e$) and the fidelity $\mathcal{F} =
N^{-1}\sum_i \vec{S}_i \cdot \vec{S'}_i$, i.e. defined as the
average dot product of spins on each lattice point of two
configurations $\vec{S}(\vec{r})$ and $\vec{S'}(\vec{r})$. The
energy of our parametrization (Fig. 6(b)) is significantly closer to
the energy obtained by Monte Carlo (Fig. 6(a)) and its fidelity is
significantly closer to 1 than the two single vortex parametrization
(Fig. 6(c)). This is important because the energies of the single
vortex and the double vortex configurations are very close. If one
uses the naive parametrization, the double vortex could never be the
ground state.

Using the parametrization of the double vortex in Eq.\ref{ellipse},
we now apply the scaling procedure to obtain the phase diagram for
elliptically shaped particles (see Fig. 7). Since there is no good
description for the core of the double vortex yet, a core-free
system is assumed for simplicity. We estimate that the boundary will
shift to lower values of $R_a$ and $H$ by about $35\%$ when taking
the core into consideration.

\begin{figure}[h]
\includegraphics[height=.25\textheight]{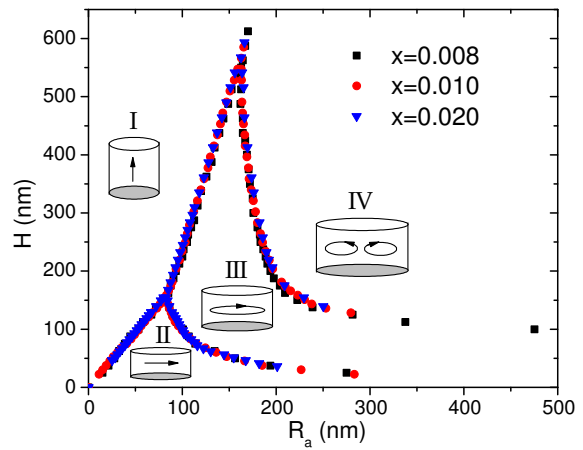}
\caption{(color online) Scaled phase diagram of an elliptically
shaped magnetic nanoparticle ($Ka^3/D=1$ and $Ja^3/D=5000$) as a
function of its semi-major axis ($R_a$) and height ($H$) with an
aspect ratio 2. The four competing phases are (I) out-of-plane
ferromagnetism, (II) in-plane ferromagnetism, (III) single vortex
state and (IV) double vortex state. The scaling exponent is
$\eta=0.55$.}
\end{figure}

As expected, the double vortex state becomes stable when both the
semi-major axis and height of the nanoparticle are increased. In the
vicinity of the phase boundary between the single vortex and the
double vortex states, the energies for the two configurations are
very close, and hence there could be a large metastable region close
this phase boundary where both states could exist in nature. This is
likely the reason why both these configurations have been observed
in experiments on the same particle\cite{ellipseNature}. Regarding
the scaling exponent, $\eta=0.55$ is again observed in this
core-free consideration, implying incomplete self-similarity.

Here we have only focused on the double vortex state. When the
system size and the aspect ratio are sufficiently large, it is
possible that multivortex states emerge. Besides such complex single
domain structures, cross-tie domain walls\cite{crosstie} could exist
in these structures as well. It would be highly interesting to know
under which condition these configurations could be stabilized.


\section{Conclusions}
In conclusion, we have extended and analyzed the hypothesis of
physical similarity put forward in Ref\onlinecite{scalecylinder}.
Regardless of shape, anisotropy, or crystal structure, we find
numerical evidence for {\em incomplete} similarity ($\eta=0.55$)
with respect to the lattice constant $a$ when a ``core-free" model
is assumed. Introducing additional small length scales, such as
core-size or an inner radius, restores {\em complete} similarity
($\eta=0.5$), since the new small length regulates the vortex core.

A three-dimensional phase diagram for the cylindrical ring structure
was obtained and a linear relationship between the height ($H_t$) at
the triple point and the inner radius ($R_i$) was found, which
offers a straightforward way to calculate the critical inner radius
above which there exists no in-plane ferromagnetic phase. A new
parametrization for double vortex configurations was proposed. This
configuration was found to be the ground state when both the radius
and height of the elliptically shaped magnetic particle are large.
Finally, a new phase diagram for elliptical nanoparticles including
a double vortex phase was determined.

\textsc{\textbf{Acknowledgement:}} We would like to thank Yaqi Tao,
Denis Koltsov, Ilya Krivorotov and Jose d'Albuquerque e Castro for
useful discussions concerning this topic as well closely related
problems. The computing facility is generously provided by USC
high-performance supercomputing center. We also acknowledge
financial support by the Department of Energy under grant
DE-FG02-05ER46240.

\bibliography{scale}        

\end{document}